\documentclass[journal]{IEEEtran}

\usepackage{amsmath, amssymb, bm}
\usepackage{cite}
\usepackage{graphicx}
\usepackage{multirow}
\usepackage{xcolor}
\usepackage{balance}
\usepackage[font={small}]{caption}
\usepackage{algorithmic}
\usepackage{url}
\usepackage{adjustbox}
\usepackage{soul}
\newcommand{\norm}[1]{\left\lVert#1\right\rVert}

\ifodd 1 
\newcommand{\rev}[1]{#1}
\newcommand{\revm}[1]{#1}
\newcommand{\final}[1]{#1}
\else
\newcommand{\rev}[1]{#1}
\newcommand{\revm}[1]{#1}
\fi

\graphicspath{{./figures/}}

\begin{document}

\title{A Multi-Channel Ratio-of-Ratios Method for Noncontact Hand Video Based SpO$_2$ Monitoring Using Smartphone Cameras} %

\author{
Xin~Tian,~\IEEEmembership{\revm{Graduate} Student Member,~IEEE,} Chau-Wai~Wong,~\IEEEmembership{Member,~IEEE,}
Sushant~M.~Ranadive,
~and~Min~Wu,~\IEEEmembership{Fellow,~IEEE}

\thanks{Manuscript received May 30, 2021; revised January 24, 2022; accepted January 29, 2022. This work was supported \final{in part} by NSF ECCS under Grants 2030502 and 2030430, and a UM Venture COVID Challenge award.}

\thanks{X. Tian and M. Wu are with the Department of Electrical and Computer Engineering, University of Maryland, College Park, MD 20742, USA (e-mail: \{xtian17, minwu\}@umd.edu).}

\thanks{C.-W. Wong is with the \revm{Department of Electrical and Computer Engineering, the Forensic Sciences Cluster, and the Secure Computing Institute,} NC State University, Raleigh, NC 27695, USA (email: chauwai.wong@ncsu.edu).}

\thanks{S. M. Ranadive is with the Department of Kinesiology, School of Public Health, University of Maryland, College Park, MD 20742, USA (e-mail: ranadive@umd.edu).}

\thanks{2022 IEEE Journal of Selected Topics in Signal Processing. Personal use is permitted, but republication/redistribution requires IEEE permission.}

}

\maketitle

\begin{abstract}
Blood oxygen saturation (SpO$_2$) is an important indicator for pulmonary and respiratory functionalities. Clinical findings on COVID-19 show that many patients had dangerously low blood oxygen levels not long before conditions worsened. It is therefore recommended, especially for the vulnerable population, to regularly monitor the blood oxygen level for precaution. Recent works have investigated how ubiquitous smartphone cameras can be used to infer SpO$_2$. Most of these works are contact-based, requiring users to cover a phone's camera and its nearby light source with a finger to capture reemitted light from the illuminated tissue. Contact-based methods may lead to skin irritation and sanitary concerns, especially during a pandemic. In this paper, we propose a noncontact method for SpO$_2$ monitoring using hand videos acquired by smartphones. Considering the optical broadband nature of the red~(R), green~(G), and blue~(B) color channels of the smartphone cameras, we exploit all three channels of RGB sensing to distill the SpO$_2$ information beyond the traditional ratio-of-ratios (RoR) method that uses only two wavelengths. To further facilitate an accurate SpO$_2$ prediction, we design adaptive narrow bandpass filters based on accurately estimated heart rate to obtain the most cardiac-related AC component for each color channel. Experimental results show that our proposed blood oxygen estimation method can reach a mean absolute error of 1.26\% when a pulse oximeter is used as a reference, outperforming the traditional RoR method by 25\%.
\end{abstract}


\section{Introduction}
\label{sec:intro}

\IEEEPARstart{P}{eripheral} blood oxygen saturation (SpO$_2$) shows the ratio of oxygenated hemoglobin to total hemoglobin in the blood, which serves as a vital health signal for the operational functions of organs and tissues~\cite{simon2008role}. It has become increasingly important in the COVID-19 pandemic, where many patients have experienced ``silent hypoxia," a low level of SpO$_2$ even before obvious breathing difficulty is observed~\cite{shenoy2020considerations, Couzin-Frankel455, tobin2020covid}. The vulnerable population with a high possibility of infection is recommended to monitor their oxygen status continuously for early COVID-19 detection~\cite{shenoy2020considerations, teo2020early}. 

Pulse oximeters have been widely used for SpO$_2$ measurement at home and in hospitals in the form of a finger-clip~\cite{webster1997design, severinghaus2007takuo}, which adopts the ratio-of-ratios (RoR) principle. The RoR principle is based on the different optical absorption rates of the oxygenated hemoglobin (HbO$_2$) and deoxygenated hemoglobin (Hb) at $660$~nm (red) and $940$~nm (infrared) wavelengths. By illuminating red and infrared lights on the fingertip, the intensity of the transmitted light on the receiver end of the pulse oximeter contains pulsatile information to derive SpO$_2$. The gold standard for measuring the blood oxygen saturation is blood gas analysis, which is invasive and painful and requires well-trained healthcare providers to perform the test. In contrast, the pulse oximeter is noninvasive and provides readings in nearly real time, and is therefore more tolerated and convenient for daily use.
The pulse oximeter is known to have a deviation of $\pm2\%$ from the gold standard when the blood oxygen saturation is in the range of $70\%$ to $99\%$~\cite{pluddemann2011pulse}, which is well-known and accepted in the clinical use.

With the ubiquity of smartphones and the growing market of smart fitness devices, the RoR principle has been applied to new nonclinical settings for SpO$_2$ measurement. Apple Watch Series 6 has blood oxygen measurement functionality, and it requires skin contact with the watch neither too tight nor too loose for the best results~\cite{applespo2}. 
The recent scientific literature also explored methods for SpO$_2$ estimation using a smartphone. 
These methods require a user to use his/her fingertip to cover an optical sensor and a nearby light source to capture the reemitted light from the illuminated tissue~\cite{scully2011physiological, lu2015prototype, ding2018measuring, bui2020smartphone, tayfur2019reliability}.
The aforementioned SpO$_2$ estimation methods based on smartphone and smartwatch are contact-based. {It can easily spread germs or diseases if shared among multiple persons,} and may irritate sensitive skin or cause a sense of burning from the heat built up if {a} fingertip is in contact with the flashlight for an extended period of time.

Researchers have recently investigated measuring the saturation of blood oxygen by means of contactless techniques~\cite{kong2013non, van2016new, guazzi2015non, van2019data}. These methods typically acquire a user’s face video under ambient light with CCD cameras to estimate SpO$_2$ from pulsatile information of monochromatic wavelengths. Shao~{et~al.}~\cite{shao2015noncontact} also use a facial video-based method to monitor SpO$_2$ that is implemented using a CMOS camera with a light source alternating at two different wavelengths. Tsai~{et~al.}~\cite{tsai2016no} acquire hand images with CCD cameras under two monochromatic lights to analyze SpO$_2$ from the reflective intensity of the shallow skin tissue. These contactless methods can provide alternatives of contact-based SpO$_2$ measurements for individuals with finger injuries or nail polish~\cite{cote1988effect, yont2014effect}, for whom the traditional pulse oximeters may be inaccurate. However, the setups used in the abovementioned studies are expensive and not common for daily use.  

More economical camera devices, low-cost webcams~\cite{rosa2019noncontact, casalino2020mhealth, bal2015non, tarassenko2014non} and smartphones~\cite{sun2021robust} have also been applied for contactless SpO$_2$ estimation. Unlike pulse oximeters, all three color channels in these economical RGB cameras capture a wide range of wavelengths from the ambient light. Most works using RGB cameras~\cite{rosa2019noncontact, casalino2020mhealth, bal2015non, tarassenko2014non, sun2021robust} directly apply the conventional narrowband red--infrared RoR principle to the red and blue (or green) channels of RGB videos without addressing the broadband nature of the color channels.

In this paper, we propose a multi-channel RoR method for noncontact SpO$_2$ monitoring using hand videos captured by smartphone cameras under ambient light. The contributions of our work include:

\begin{itemize}

\item 
We exploit all three RGB channels to extract features for SpO$_2$ prediction, instead of limiting to two wavelengths/color channels as in traditional RoR methods. Efficiently utilizing three channels is nontrivial and is one of the key research issues we shall address in this paper. We will take into consideration the underlying optophysiological model given the smartphone camera as the remote sensor and the ambient light environment. 
Our proposed multi-channel RoR based method can achieve a mean absolute error of $1.26\%$ in SpO$_2$ estimation with the pulse oximeter as the reference, which is $25\%$ lower than that of the traditional RoR model.

\item 
We filter the RGB signals with a {narrow adaptive bandpass (ABP) filter} centered at an {accurately} estimated heart rate (HR) to obtain the most relevant cardiovascular-related \rev{AC component. We systematically verify} the important roles of both the narrow ABP filter and the accurate HR tracking for accurate SpO$_2$ monitoring.

\item 
We investigate the more favorable scenario for data collection by using \final{the} hand as the signal source for the period that \final{the} face-covering mandate is on. It has a few practical advantages compared to facial videos, including being applicable with masks on, less privacy concern, and potentially being more tolerant to different skin tones than the face. We analyze the impact of the sides of the hand and skin tones on the SpO$_2$ estimation \rev{performance.}

\end{itemize}

\section{Background and Related Work}
\noindent In this section, we review the ratio-of-ratios (RoR) model for SpO$_2$ measurement. Consider a light source with spectral distribution $I(\lambda)$ illuminating the skin, and a remote color camera with spectral responsivity $r({\lambda})$ recording a video. The light from the source travels through the tissue and part of the light in the tissue is reemitted to be received by the color camera. During each cardiac cycle, the heart muscle contracts and relaxes, so that the blood is \revm{pumped} to the body and travels back to the heart. During this process, the blood volume increases and decreases in the arterial vessels, causing increased and decreased light absorption~\cite{sun2015photoplethysmography}. According to the skin-reflection model~\cite{wang2016algorithmic, szeliski2010computer}, the color camera will receive the specularly reflected light from the skin surface and the diffusely reemitted light from the tissue--light interaction that contains the cardiac-related pulsatile information. Based on the verified assumption proposed in~\cite{guazzi2015non} that the specular reflection components can be ignored if the movement is minimized, the camera sensor response at time $t$ \revm{is:}

\begin{equation}
\label{sc in general}
    \mathcal{S}_c(t) = \int_{\Lambda_c} {I}(\lambda) \cdot e^{-\mu_d(\lambda,t)} \cdot r_c(\lambda)~  d\lambda .
\end{equation}
where the $\lambda$ is the wavelength, the integral range $\Lambda_c$ captures the responsive wavelength band of channel $c$ of the camera, $I(\lambda)$ is the spectral intensity of the light source, $\mu_d(\lambda,t)$ is the diffusion coefficient, and $r_c(\lambda)$ is the sensor response of channel $c$ of the camera.

According to the {Beer-Lambert’s law~\cite{van2016new}}, the diffusion coefficient $\mu_d(\lambda, t)$ can be expanded into:
\begin{equation}
    \begin{aligned}
     \mu_d(\lambda,t) & = \varepsilon_\text{t}(\lambda) \, C_\text{t} \, l_\text{t} \\&+  [\varepsilon_{\text{Hb}}(\lambda)C_{\text{Hb}} + \varepsilon_{\text{HbO}_2}(\lambda)C_{\text{HbO}_2}] \, l(t),
    \end{aligned}
\end{equation}
where $\varepsilon_{\text{Hb}}$, $\varepsilon_{\text{HbO}_2}$, and $\varepsilon_\text{t}$ are the extinction coefficients of arterial deoxyhemoglobin, arterial oxyhemoglobin, and other tissues including the venous blood vessel, respectively; $C_\text{t}$, $C_{\text{Hb}}$, and $C_{\text{HbO}_2}$ are the concentrations of the corresponding substances; $l_\text{t}$ is the path length that the light travels in the tissue and is assumed to be time-invariant; $l(t)$ is the path length that the light travels in the arterial blood vessels. Note that $l(t)$ is time-varying because the arteries will dilate with increased blood during systole compared to diastole.

The integral range $\Lambda_c$ can be simplified to a single value $\lambda_i$ when the camera is monochromatic, incoming light is filtered by a narrowband optical filter, or alternatively, the light source is a narrowband LED. Then the response of the camera sensor~in~\eqref{sc in general} can be written as:
\begin{equation}
\label{eq. narrowband signal}
    \begin{aligned}
    \mathcal{S}_c(t) =   &{I}(\lambda_i) \cdot e^{-\varepsilon_\text{t}(\lambda_i)C_\text{t}l_\text{t}} \cdot r_c(\lambda_i) \\& \cdot e^{-{[\varepsilon_{\text{Hb}}(\lambda_i)C_{\text{Hb}}+\varepsilon_{\text{HbO}_2}(\lambda_i)C_{\text{HbO}_2}]\, l(t)}}.
    \end{aligned}
\end{equation}
Let $\Delta l = l_\text{max} - l_\text{min}$ denote the difference of the light path of the pulsatile arterial blood between diastole when $l(t) = l_\text{min}$ and systole when $l(t) = l_\text{max}$. The log-ratio of the response of the $c$th channel of the camera sensor during diastole and systole can then be written as:

\begin{subequations}
\label{eq.ratio in narrwoband setting}
\begin{alignat}{2}
    {R}(\lambda_i) & =  \log \left( \frac{\mathcal{S}_c|_{l(t)=l_\text{min}}}{\mathcal{S}_c|_{l(t)=l_\text{max}}} \right)\\&=  [\varepsilon_{\text{Hb}}(\lambda_i)C_{\text{Hb}}+\varepsilon_{\text{HbO}_2}(\lambda_i)C_{\text{HbO}_2}]\, \Delta l.
\end{alignat}
\end{subequations}
For two different wavelengths $\lambda_1$ and $\lambda_2$, the ratio of ratios (RoR) can then be defined as:
\begin{equation}
\label{ror}
    \text{RoR}(\lambda_1, \lambda_2)\!=\!\frac{{R}(\lambda_1)}{{R}(\lambda_2)}\!=\!\frac{\varepsilon_{\text{Hb}}(\lambda_1)C_{\text{Hb}}\!+\!\varepsilon_{\text{HbO}_2}(\lambda_1)C_{\text{HbO}_2}}{\varepsilon_{\text{Hb}}(\lambda_2)C_{\text{Hb}}\!+\!\varepsilon_{\text{HbO}_2}(\lambda_2)C_{\text{HbO}_2}}.
\end{equation}

\noindent Since $\text{SpO}_2 = C_{\text{HbO}_2} \big/ \left( C_{\text{HbO}_2}+C_{\text{Hb}} \right)$, the relation between RoR and SpO$_2$ can be written as:
\begin{subequations}
\begin{alignat}{2}
    \text{SpO}_2 &\!=\! \frac{\varepsilon_{\text{Hb}}(\lambda_1)\!-\! \varepsilon_{\text{Hb}}(\lambda_2)\!\cdot\!\text{RoR}}{\varepsilon_{\text{Hb}}(\lambda_1)\!-\!\varepsilon_{\text{HbO}_2}(\lambda_1)\!+\![\varepsilon_{\text{HbO}_2}(\lambda_2)\!-\!\varepsilon_{\text{Hb}}(\lambda_2)]\!\cdot\! \text{RoR}}\\
    & \approx \alpha \!\cdot\! \text{RoR} \!+\! \beta.
    \label{linear relation spo2 and rr}
\end{alignat}
\end{subequations}
where the linear approximation can be obtained by a Taylor expansion.

The linear RoR model in~\eqref{linear relation spo2 and rr} has been applied under different SpO$_2$ measurement scenarios. For pulse oximeters, $\lambda_1 = 660$~nm and $\lambda_2 = 940$~nm are used to leverage the optical absorption difference of Hb and HbO$_2$ at the two wavelengths. In some prior art using narrowband light sources or monochromatic camera sensors~\cite{kong2013non, shao2015noncontact} for contactless SpO$_2$ monitoring, different combinations of $(\lambda_1, \lambda_2)$ have been explored. In the prior art using consumer-grade RGB cameras~\cite{rosa2019noncontact, casalino2020mhealth, bal2015non, tarassenko2014non, sun2021robust}, only two out of the three available RGB channels were used for the linear RoR model. 

Among the abovementioned SpO$_2$ estimation methods using consumer-grade RGB cameras, the SpO$_2$ data collected in~\cite{rosa2019noncontact, casalino2020mhealth} only cover a small dynamic range (mostly above $95\%$), which is not very meaningful. Bal~{et~al.}~\cite{bal2015non} and Tarassenko~{et~al.}~\cite{tarassenko2014non} show a fitted linear relation between RoR and SpO$_2$ for data that last for merely several minutes. These limitations can be attributed to that, unlike the signals captured in the narrowband setting that is modeled precisely by~\eqref{eq. narrowband signal} and~\eqref{eq.ratio in narrwoband setting}, all three RGB color channels capture a wide range of wavelengths from the ambient light, as is described in~\eqref{sc in general}. The aggregation of the broad range of wavelengths lowers the optical difference between Hb and HbO$_2$ and makes it less optically selective than narrowband sensors used in oximeters. 
To address this issue, we disentangle the aggregation 
through a careful combination of the pulsatile signals from all three channels of RGB videos to efficiently distill the SpO$_2$ information.

\section{Proposed Multi-Channel RoR Method}
\label{sec:framework}

\begin{figure}
\centering
\includegraphics[width=3.5in]{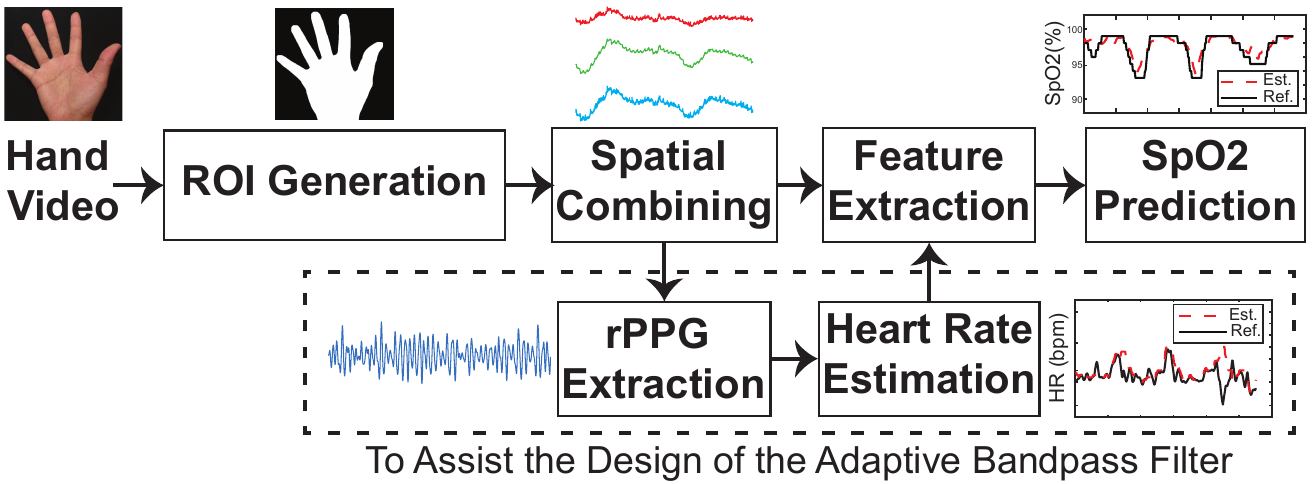}
\caption{System illustration for the SpO$_2$ prediction using the smartphone captured hand videos. The pixels from the hand region are utilized for prediction, and an rPPG signal is extracted for heart rate (HR) estimation. Multi-channel RoR features are derived from the spatially combined RGB signals with the help of the HR-guided filters. The extracted features are then used for SpO$_2$ prediction.}
\label{fig::flowchart}
\end{figure} 

Fig.~\ref{fig::flowchart} illustrates the proposed procedure for the SpO$_2$ estimation from the smartphone captured hand videos. First, the hand is detected as the region of interest (ROI) for each frame. Second, the spatial average from the ROI is calculated to obtain three time-varying signals of RGB channels. The averaged RGB signals are extracted for two purposes: i) to estimate the heart rate (HR), and ii) to acquire the filtered cardio-related AC components using an HR-based adaptive bandpass filter. Third, the ratio between the AC and the DC components for each color channel and the pairwise ratios of the resulting three ratios are computed as the features for a regression model where SpO$_2$ is treated as the label. The details of each step are provided as follows.

\subsection{ROI Generation via Thresholding and Spatial Combining}
\label{subsec:roi}

To facilitate the data collection with good quality, we use a rectangle to enclose the target hand region. We use an interactive user interface for this step, which can be replaced by an automated hand detection algorithm~\cite{urooj2018analysis} if desired. The pixels in this region are converted from the RGB color space to the YCrCb color space, and the Cr channel is used~\cite{chai1999face} to determine a threshold that best differentiates the skin pixels from the background using the Otsu algorithm~\cite{otsu1979threshold}. We apply morphological erosion and dilation operations with a median filter to exclude noise pixels outside the region of the binary hand mask. The final hand-shaped mask is considered as the ROI, and an example is shown in the second picture in Fig.~\ref{fig::flowchart}. \rev{We conducted the ROI segmentation independently for each video frame to capture any unintended tiny hand movement rather than using a fixed hand mask for all frames.} For all $n$ frames in the video, we calculate the spatial average values of the red, green, and blue channels in the ROI and denote them as $\mathbf{\bar{r}}, \mathbf{\bar{g}}, \mathbf{\bar{b}} \in \mathbb{R}^{1\times n}$, and arrange them into a matrix $\mathbf{A} = [\mathbf{\bar{r}}; \mathbf{\bar{g}}; \mathbf{\bar{b}}] \in \mathbb{R}^{3\times n}$.

\subsection{rPPG Extraction and HR Estimation}
\label{subsec:hr-esti}

In a typical RoR method, after matrix {$\mathbf{A}$} 
is calculated, the AC component for each channel of {$\mathbf{A}$} is quantified by either the standard deviation~\cite{scully2011physiological} or the peak-to-valley amplitude~\cite{shao2015noncontact}. Since the signal-to-noise ratio~(SNR) is lower for the video captured by a smartphone in a contactless manner, we propose to use an adaptive bandpass filter centered at the HR frequency to clean the RGB channel signals so as to extract the AC components more precisely.

\begin{figure}
\centering
\includegraphics[width=3.5in]{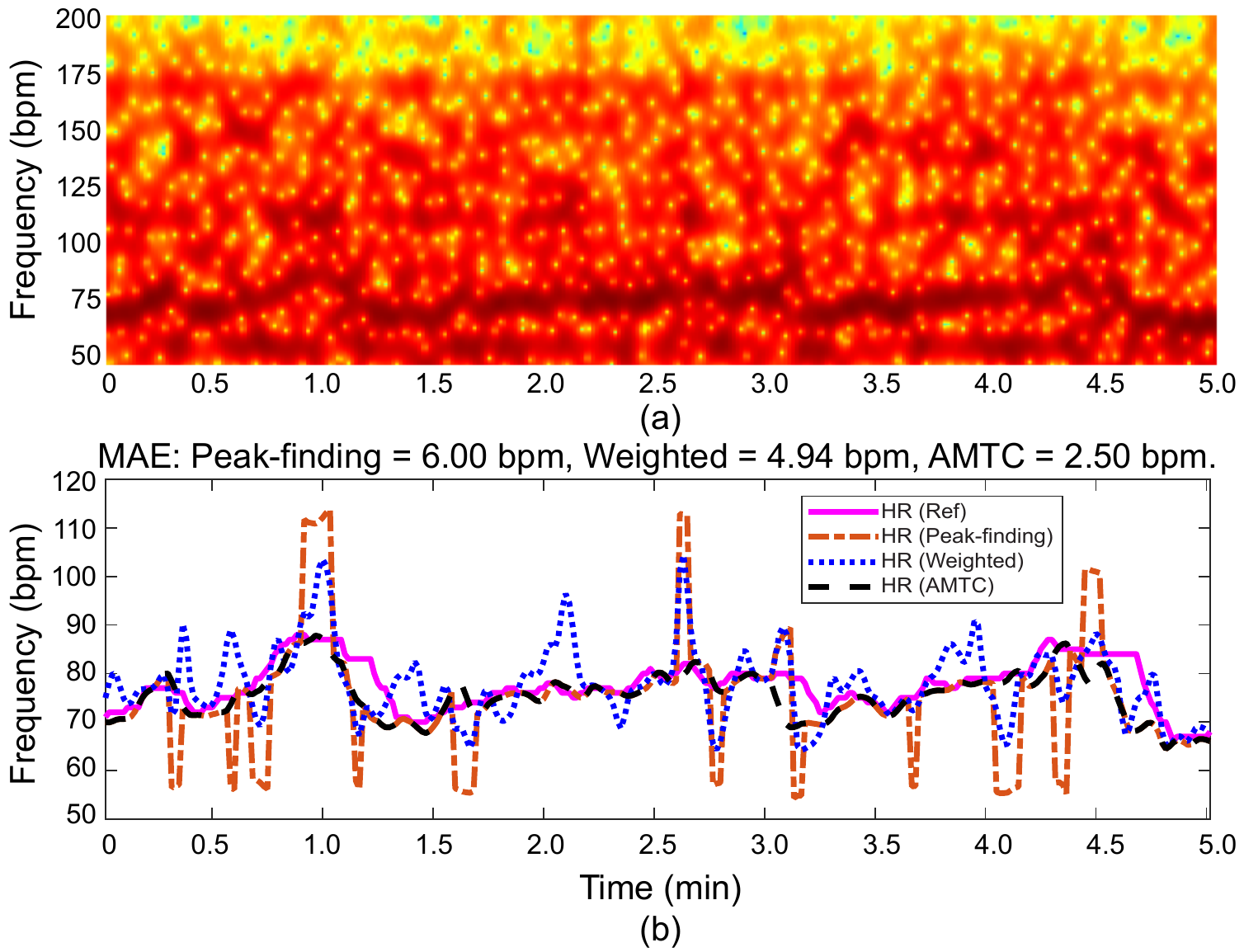}
\caption{(a) Spectrogram of an rPPG signal. (b) Reference HR signal and HR signals estimated by the peak-finding method, the weighted energy frequency estimation algorithm, and AMTC algorithm, respectively. The mean absolute error (MAE) of the HR estimation algorithms are 6.00, 4.94, and 2.50 bpm, respectively.}
\label{fig::hr result}
\end{figure}

The HR can be measured contact-free by capturing the pulse-induced subtle color variations of the skin. The pulse signal, referred to as remote photoplethysmogram (rPPG), can be obtained \final{by} applying the plane-orthogonal-to-skin (POS) algorithm~\cite{wang2016algorithmic}, which defines a plane orthogonal to the skin tone in the RGB space for robust rPPG extraction. The HR is then tracked from the rPPG signal via a state-of-the-art adaptive multi-trace carving (AMTC)~\cite{zhu2020adaptive} algorithm that tracks the HR from the spectrogram of rPPG by dynamic programming and adaptive trace compensation.

To study the role of accurate HR tracking for feature extraction, we also implemented a peak-finding method and a weighted energy method for frequency estimation~\cite{hajj2012instantaneous} to compare with AMTC. The peak-finding method takes the peaks of the squared magnitude of the Fourier transform of rPPG as the estimated HR values, which was used in~\cite{van2016new} and~\cite{guazzi2015non}. The weighted energy method finds the heart rate by weighing the frequency bins in the corresponding frame of the spectrogram of rPPG. Compared to the peak-finding method, the weighted energy method is more robust to outliers in frequency. Fig.~\ref{fig::hr result} illustrates an example of the HR estimation results by the peak-finding method, the weighted energy algorithm, and AMTC, respectively.

\subsection{Feature Extraction}
\label{subsec:feature-extraction}
We use a processing window of {10 seconds} with a step size of 1 second to segment the whole video into $L$ windows. Within each window, the DC and AC components of the RGB channels are calculated to build a feature vector~$\mathbf{f}$.

\noindent\textbf{DC component} We use a second-order lowpass Butterworth filter with a cutoff frequency \revm{of} 0.1 Hz. The DC component is estimated using the median of the lowpass filtered signal of each window.

\noindent\textbf{AC component} The estimated heart rate values from Section~\ref{subsec:hr-esti} are used as the center frequencies for the adaptive bandpass (ABP) filters to extract the AC components of the RGB channels, which eliminates all frequency components that are unrelated to the cardiac pulse. We adopt an 8th-order Butterworth bandpass filter with $\pm0.1$~Hz~($\pm$6~bpm) bandwidth, centering at the estimated HR of the current window. The magnitude of the AC component is estimated using the average peak-to-valley amplitudes of the filtered signals within the current processing window. 

We define the normalized AC components at the $i$th window as ${R}(i,c)=\frac{\text{AC}(i,c)}{\text{DC}(i,c)}$, where $c \in {\{r, g, b\}}$ represents color channel and $i\in\{1,2,...,L\}$. We define the multi-channel ratio-of-ratios based feature vector of the $i$th window as ${\mathbf{f}}_i~=~[{R}(i,r),{R}(i,g),{R}(i,b),\frac{{R}(i,r)}{{R}(i,g)},\frac{{R}(i,r)}{{R}(i,b)},\frac{{R}(i,g)}{{R}(i,b)}]~\in~\mathbb{R}^{1 \times 6}$.

\subsection{Regression and Prediction}

In this first proof-of-concept work leveraging multi-channel ratio features, we use linear regression (LR) and support vector regression (SVR) to learn the mapping between the features and the SpO$_2$ level. Since LR captures only the linear relationship, it has limited learning capability and will serve as a baseline. The objective function is ${\underset{\mathbf{w}}{\min}\norm{\mathbf{y} \!-\!  \mathbf{F} \mathbf{w}}^2 \!+\! \lambda\norm{\mathbf{w}}_2^2}$, where $\mathbf{y}=[y_1,...,y_L]^\text{T}\in\mathbb{R}^{L\times1}$ contains the target SpO$_2$ values, 
${\mathbf{F}=[\mathbf{f}_1;...;\mathbf{f}_L]\in\mathbb{R}^{L\times6}}$ is the feature/data matrix derived from the input, and
$\mathbf{w}\in\mathbb{R}^{6\times1}$ contains the weights. An $\ell_2$-regularization term is added to the objective function to avoid rank deficiency caused by the collinearity among features. To select the optimal regularization parameter $\lambda$, we use a 5-fold cross-validation.
In addition to LR, we use the SVR to better capture the nonlinearity of the features. The Libsvm library~\cite{chang2011libsvm} is used for training the $\epsilon$-SVR, ${\underset{\mathbf{w}, b}{\min}\,\frac{1}{2}\norm{\mathbf{w}}^2\! +\!  C \sum_{i=1}^L \mathcal{L}_{\epsilon} \left( y_i, \mathbf{w}^\text{T}\boldsymbol{\phi}(\mathbf{f}_i)\!+\! b \right)}$, where $\mathcal{L}_{\epsilon}$ is the linear $\epsilon$-insensitive loss function~\cite{theodoridis2015machine}. Our implementation uses the radial basis function (RBF) kernel to capture the nonlinearity. The hyperparameters, including the penalty cost $C$ and the kernel parameter $\gamma$ of kernel function ${K(\mathbf{f}_i,\mathbf{f}_j)=\boldsymbol{\phi}(\mathbf{f}_i)^\text{T}\boldsymbol{\phi}(\mathbf{f}_j) = \exp \left({-\gamma\norm{\mathbf{f}_i-\mathbf{f}_j}^2} \right)}$, are selected via a grid search over a 5-fold cross-validation loss.

Once an estimated weight vector $\hat{\mathbf{w}}$ is learned from the linear or support vector regression, $\hat{\mathbf{w}}$ is then used to predict a preliminary SpO$_2$ signal. 
Finally, a 10-second moving average window is applied to smooth out the preliminarily predicted signal to obtain the final predicted SpO$_2$ signal.

\section{Experimental Results}
\label{sec:expr}

\subsection{Data Collection}
\begin{figure}[!t]
\centering
\includegraphics[width=2in]{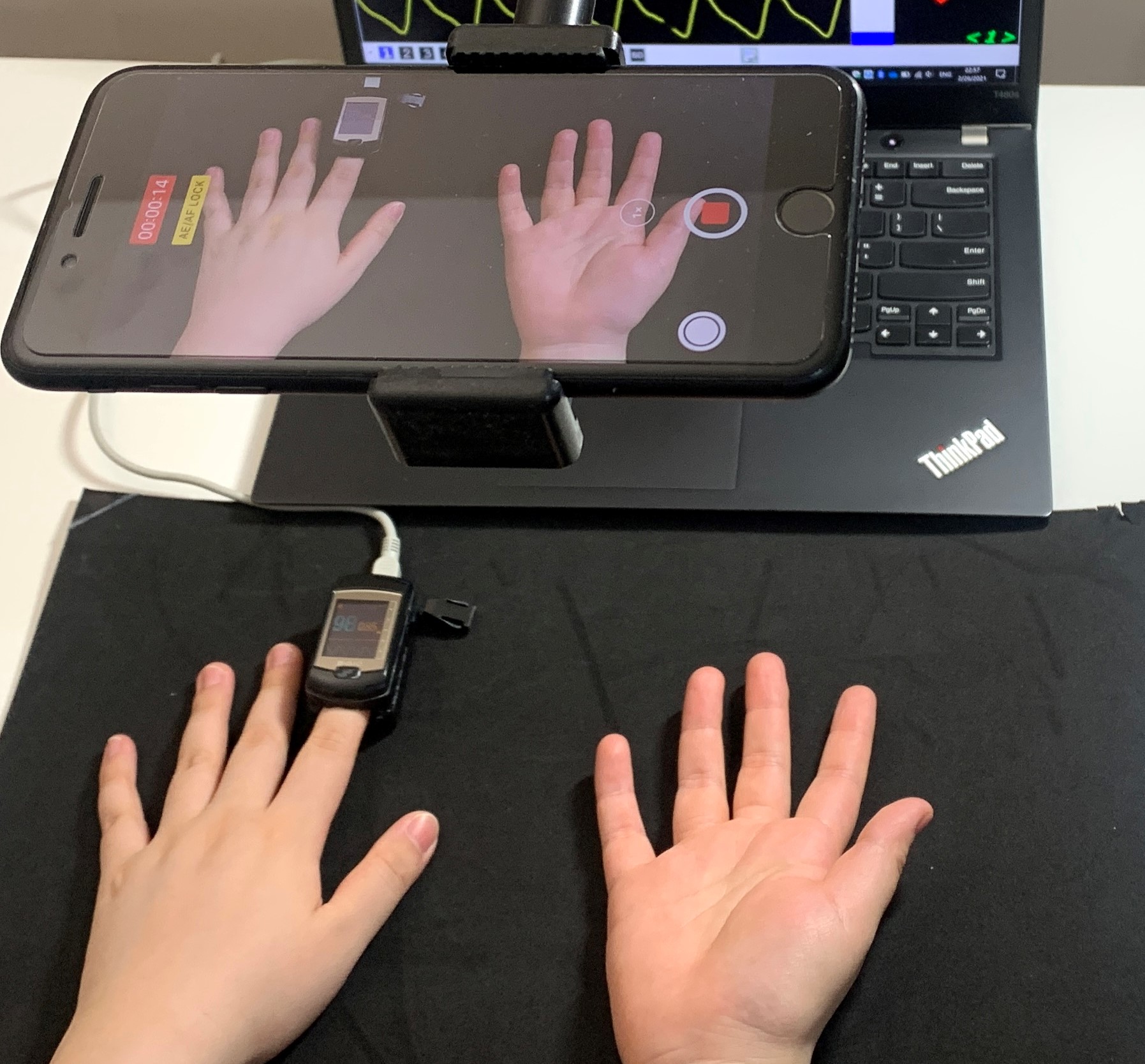}
\caption{Experimental setup for data collection of hand videos and reference signals using an oximeter. The left index finger was placed in a CMS-50E pulse oximeter to record the reference HR and SpO$_2$ signals. The smartphone camera is recording the video of both hands.}
\label{fig::setup}
\end{figure}

\begin{figure*}[!t]
\centering
\includegraphics[width = 6.8in]{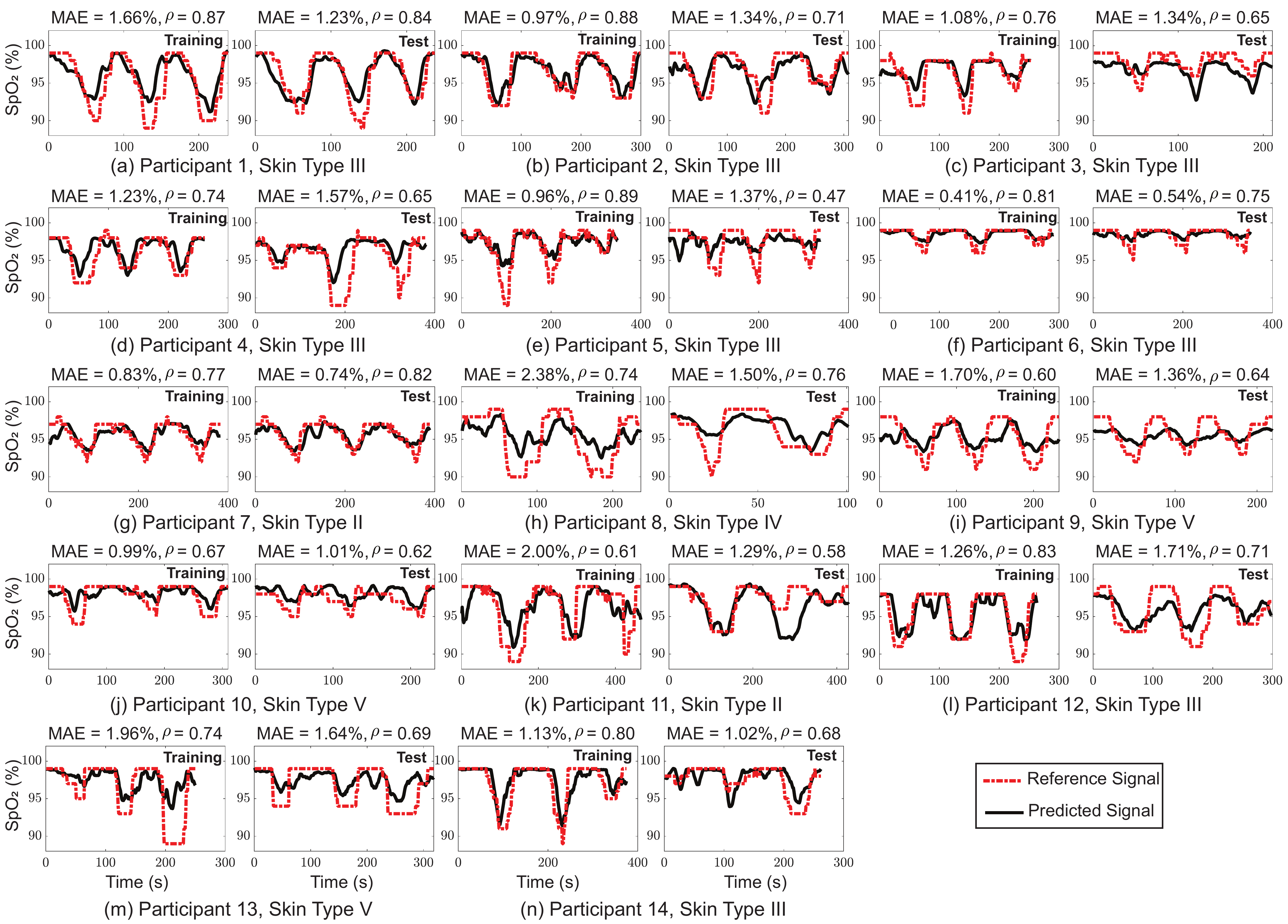}
\caption{Predicted SpO$_2$ signals for all participants using SVR when the palm is facing the camera, i.e., the palm-up scenario. Prediction results of training and testing sessions are shown for each participant with reference SpO$_2$ in red dash lines and predicted SpO$_2$ in solid black lines. The higher the correlation $\rho$ and the lower the MAE, the better the predicted SpO$_2$ captures the trend of the reference signal.}
\label{fig::fig1}
\end{figure*} 

Fourteen volunteers, including eight females and six males, were enrolled in our study under protocol \#1376735 approved by the University of Maryland Institutional Review Board (IRB), with age range between $21$ and $30$, and Fitzpatrick skin types II--V~\rev{\cite{skin-types}}. There are two, eight, one, and three participants having skin types II, III, IV, and V, respectively. None of the participants had any known cardiovascular or respiratory diseases. During the data collection, participants were asked to hold their breath to induce a wide dynamic range of SpO$_2$ levels. The typical SpO$_2$ range for a healthy person is from $95\%~\text{to}~100\%$. By holding breath, the SpO$_2$ level can drop below $90\%$. Once the participant resumes normal breathing, the SpO$_2$ will return to the level before the breath-holding.

Each participant was recorded for two sessions. During the recording, the participant sat comfortably in an upright position and put both hands on a clean dark foam sheet placed on a table. As shown in Fig.~\ref{fig::setup}, the palm side of the right hand and the back side of the left hand were facing the camera. These two hand-video capturing positions are defined as \emph{palm up~(PU)} and \emph{palm down~(PD)}, respectively. \rev{The participant was asked to place his/her hands still on the table to avoid hand motion.} Simultaneously, a Contec CMS-50E pulse oximeter was clipped to the left index finger to measure the participant's SpO$_2$ level. As we have reviewed earlier, the oximeter is adopted clinically as to be within a $\pm2\%$ deviation from the invasive, gold standard for SpO$_2$ \cite{pluddemann2011pulse}, so we use the oximeter measurement results as the reference in our experiments. An iPhone~7 Plus camera was \rev{fixed by a smartphone stand mounted on a tripod} for video recording. The video started 30 seconds before the oximeter \revm{started}, and stopped immediately after the oximeter \revm{ended} to allow for proper time synchronization. {The participants were asked to hold their breath for generally 30--40 seconds to lower the SpO$_2$ level, as long as they were comfortable and able to do so. Then the participants resumed normal breathing for generally 30--40 seconds until they recovered and felt ready for the next breath-holding. The recovery period was long enough for the participants' SpO$_2$ to return to the levels before the breath-holding.}
The aforementioned process is defined as one breath-holding cycle. In each session, the breath-holding cycles were repeated three times. After the first session, the participants were asked to relax for at least $15$ minutes before attending the second session for data collection. From our data collection protocol using breath-holding, we were able to obtain the SpO$_2$ measurements \revm{ranging from $89\%$ to $99\%$.}

The total length of recording time for all fourteen participants is 138.9 minutes. \rev{In terms of each participant, the minimum duration is 103 seconds and the maximum duration is 468 seconds. The average duration is 298 seconds.} The current data size is relatively small for large-scale neural network training. This is by a large part due to the restrictions for human subject related data collection imposed during the COVID-19. The available data, however, is adequate for our principled multi-channel signal based approach to SpO$_2$ monitoring, showing a benefit of combining signal processing and biomedical knowledge and modeling with data than the primarily data-driven approach. 

\noindent \textbf{Delay Estimation of Pulse Oximeter}: 
When the CMS-50E oximeter is turned on and ready for measurement, the first reading is displayed a few seconds after the finger is inserted. This delay may be due to \final{the} oximeter's internal firmware startup and algorithmic processing. Since we need to synchronize the video and the oximeter readings using their precise starting time stamps, the delay in \final{the} oximeter can introduce misalignment errors in the reference data that we use to train the regression model.
To avoid the misalignment, we first estimate the delay and then \rev{subtract it from the oximeter's internal timestamp as the corrected oximeter's timestamp}.
To \rev{estimate the internal delay}, we asked one participant to repeatedly place the left index finger, middle finger, and ring finger into the oximeter 50 times each and obtained the average delay time of 1.8s, 1.9s, and 1.7s, respectively. 
{Because the left index finger is used for reference data collection in our setup, we take 1.8s as the delay.} To further examine whether there exists any difference among the delays from the three fingers, we conducted a one-way ANOVA test. The $p$-value is $0.14$, which shows no statistically significant different delays among the three fingers.

\subsection{Performance Metrics}

The performance of the algorithm is evaluated using the mean absolute error (MAE)~\eqref{MAE equation} and Pearson's correlation coefficient $\rho$~\eqref{rho equation} given below:
\begin{subequations}
\label{eqn:rho+rmse}
\begin{alignat}{2}
    \textsc{MAE}(\mathbf{y}, \hat{\mathbf{y}}) &= \frac{1}{N}{\sum_{i=1}^N{|{y}_i - \hat{{y}}_i}|},\label{MAE equation}\\\label{rho equation}\rho(\mathbf{y}, \hat{\mathbf{y}})&=\frac{(\mathbf{y}-\bar{y})^\text{T}(\hat{\mathbf{y}}-\bar{\hat{y}})}{\norm{\mathbf{y}-\bar{y}}_2 \norm{\hat{\mathbf{y}}-\bar{\hat{y}}}_2}.
\end{alignat}
\end{subequations}
where $\mathbf{y}\!=\![y_1,...,y_N]^\text{T}$, $\hat{\mathbf{y}}\!=\![\hat{y}_1,...,\hat{y}_N]^\text{T}$, $\bar{y}$,  and $\bar{\hat{y}}$ denote the reference SpO$_2$ signal, the estimated SpO$_2$ signal, the average values of all coordinates of vectors $\mathbf{y}$ and $\hat{\mathbf{y}}$, respectively. We adopt the correlation metric to evaluate how well the trend of the SpO$_2$ signal is tracked.

\subsection{Results From Proposed Algorithm}
\label{subsec:results-main}

In this subsection, we use the training data from one participant to train the regression model for the prediction of his/her testing session recorded later. We call the aforementioned training and testing procedure the \textit{participant-specific} mode in which the models are specifically learned for each participant. We will discuss the \textit{leave-one-out} mode of the performance of the proposed algorithm in Section~\ref{subsec:leaveoneout}.

Fig.~\ref{fig::fig1} presents the learning results for all the participants using SVR for PU cases. Both training and testing sessions are shown for each participant. The SpO$_2$ curves in each session contain three dips that are resulted from breath-holding, except for participant \#8 who had a shorter session due to limited tolerance of breath-holding. For each participant, we provide the skin-tone information in the subplot and show the accuracy indicators, MAE and $\rho$, for SpO$_2$ prediction. In all training sessions, MAE is below $2.4\%$ and $\rho$ is above $0.6$. From this observation, we find that all the predicted SpO$_2$ signals in the training sessions are closely following the reference \rev{signals' trends, despite the exact value differences between the predicted and the reference signals, such as the differences around the last dip for participant~\#13.} Furthermore, all testing MAE values are within $1.8\%$, suggesting that those trained models adapt well to the testing data. While there are a few cycles that the predicted signal does not fully follow the reference signal, such as the second dip for participant \#4 and participant \#11, the trends are consistent.

\begin{table}[!t]
\caption{Performance of the proposed method. Results using linear regression (LR) and support vector regression (SVR) for both sides of the hand are quantified in terms of the sample mean and sample standard deviation (in parentheses).}
\label{table:perform summary}
\centering
\hspace{-1.5mm}
\resizebox{3.4in}{!}
{\begin{tabular}{llccccc}
\hline 
 & & \multicolumn{2}{c}{Training}  &  & \multicolumn{2}{c}{Testing} \\ \cline{3-4} \cline{6-7}&  & \multicolumn{1}{c}{MAE} & \multicolumn{1}{c}{Correlation $\rho$} &  & \multicolumn{1}{c}{MAE} & \multicolumn{1}{c}{Correlation $\rho$} \\ \hline \hline
 
\multirow{4}{*}{\textbf{LR}}  
& \multirow{2}{*}{PU} &1.69\%  & 0.63  &  & 1.52\%  & 0.62  
\\
&&($\pm$0.57\%)&($\pm$0.16)&&($\pm$0.54\%)&($\pm$0.11) \\

& \multirow{2}{*}{PD} & 1.74\% & 0.61 &  & 1.53\%   &0.56  \\
&&($\pm$0.76\%) & ($\pm$0.21)&&($\pm$0.53\%) &($\pm$0.21)\\
\hline

\multirow{4}{*}{\textbf{SVR}}
& \multirow{2}{*}{PU} & \textbf{1.33\% }& \textbf{0.76 } &  & \textbf{1.26\% } & \textbf{0.68 }  \\&&($\pm$0.54\%) & ($\pm$0.09) &&($\pm$0.33\%)&($\pm$0.10)\\

& \multirow{2}{*}{PD} & {1.35\% }& {0.75 }&  & {1.28\% }  & {0.65 } \\ &&($\pm$0.45\%) & ($\pm$0.09) &&($\pm$0.40\%)&($\pm$0.14)\\

\hline
\end{tabular}}
\end{table}
\begin{figure}[!t]
\centering
\hspace{-2mm}
\includegraphics[width=3.5in]{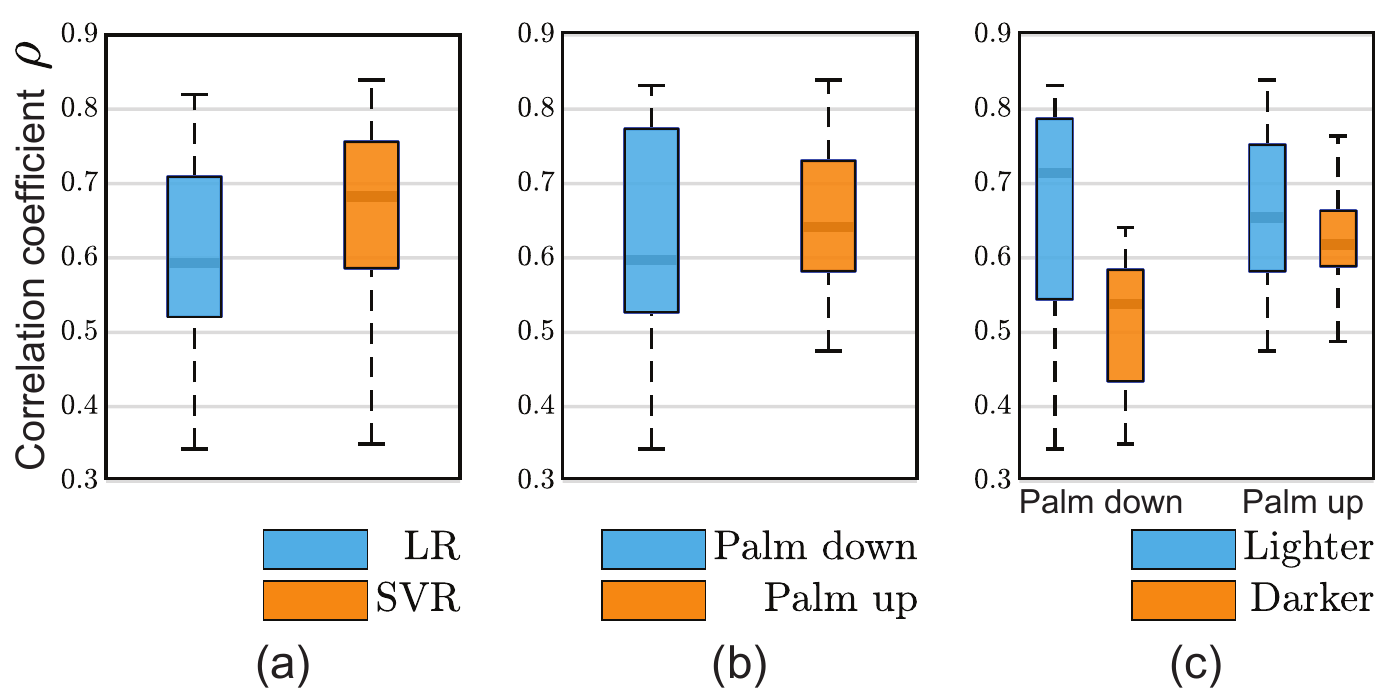}
\caption{Boxplots of testing correlation coefficient $\rho$ for all participants when grouped using different criteria. (a) Distributions contrasting linear and support vector regressions. (b) Distributions of palm-up and palm-down cases. (c) \revm{A detailed} breakdown of (b) in terms of skin-tone subgroups.}
\label{fig::boxplot}
\end{figure}

Table~\ref{table:perform summary} summarizes the training and testing SpO$_2$ estimation performance of both LR and SVR based methods for both PU and PD cases. The best performance is achieved using SVR method in the PU case. We further examine the difference between the two regression methods using boxplots in Fig.~\ref{fig::boxplot}(a) that show the distributions of the correlation $\rho$ for testing by LR and SVR, respectively. Each boxplot in Fig.~\ref{fig::boxplot}(a) contains both PU and PD cases from all participants. The results are compared in terms of the median and the interquartile range (IQR). IQR quantifies the spread of the distribution by measuring the difference between the first quartile and the third quartile. The boxplots in Fig.~\ref{fig::boxplot}(a) reveal that the SVR method outperforms LR with a higher median of 0.68 compared to 0.59 and with a narrower IQR of 0.17 compared to 0.19. This suggests that there may exist a nonlinear relationship between the extracted features and the SpO$_2$ values. 

To examine the impact of the side of a hand and the skin tone on the performance of SpO$_2$ estimation, we analyze the following two research questions: (i)~whether the side of \final{the} hand makes a difference in lighter skin (type II and III) or darker skin (type IV and V) or mixed skins (all participants), and (ii)~whether the different skin tones matter in PU or PD case. 

To answer question (i), we first focus on the distributions from PU and PD cases in Fig.~\ref{fig::boxplot}(b) with each boxplot representing the correlation $\rho$ in testing for all participants. We observe that the PU case outperforms the PD case with a higher median of 0.64 compared to 0.60 and a narrower IQR of 0.15 compared to 0.25. We then zoom into each subgroup of skin tones shown in Fig.~\ref{fig::boxplot}(c). For the lighter skin group, even though the median of PD case is 0.71, which is 9\% better than that of PU, the IQR of PD case is 0.24, which is worse than the IQR of 0.17 of PU case. This suggests that the distributions are comparable between PU and PD cases for the lighter skin group. For the darker skin group, the PU case outperforms the PD case with a higher median of 0.62 compared to 0.54 and a narrower IQR of 0.07 compared to 0.15. In summary, there is no substantial difference between PU and PD cases in the lighter skin group, whereas for the darker skin group and overall participants, the PU case is better than the PD case. 

To answer question (ii), we first focus on the left two boxplots of Fig.~\ref{fig::boxplot}(c). 
In the PD case, the median of the lighter skin group is significantly larger than that of the darker skin group by 31\%, however, the lighter skin group also has a larger IQR.  This makes it difficult to make a conclusion from the median--IQR analysis, hence we apply the $t$-test to complement our analysis. We note that the $p$-value is 0.037 $<$ 0.05, showing that there is a significant difference between these two groups.
In the PU case shown in the right half of Fig.~\ref{fig::boxplot}(c), the medians of the lighter skin group and darker skin group are 0.65 and 0.62, with IQR being 0.17 and 0.07, respectively. Thus, in our current dataset, no substantial performance difference is observed between lighter and darker skin tones in the PU case.

\subsection{Ablation Study of Proposed Pipeline}
\label{subsec:ablation-study}

\begin{figure}[!t]
\centering
\includegraphics[width=3.5in]{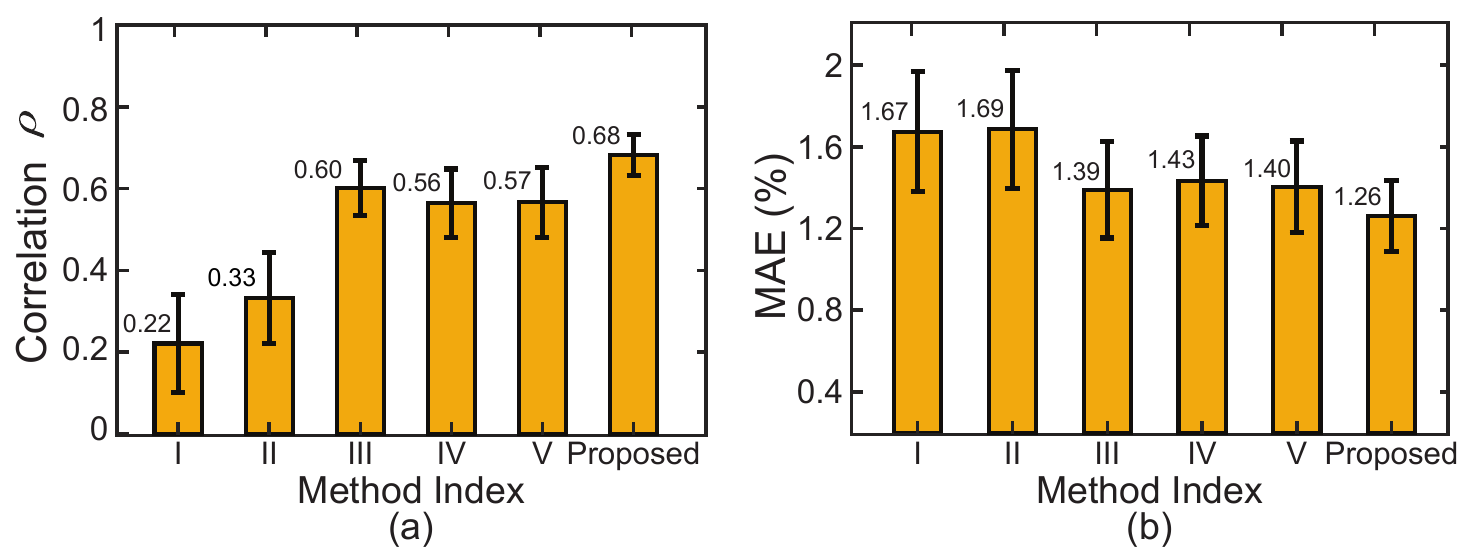}
\caption{\revm{Ablation study of the proposed method.
The bar plots are from testing sessions (SVR, PU case) of all participants. The error bars correspond to the $95\%$ confidence intervals.}}
\label{fig::barplot}
\end{figure}

In Sections~\ref{subsec:hr-esti}~and~\ref{subsec:feature-extraction}, we have proposed three key designs in our algorithm, including a)~the feature vector $\mathbf{f}$ containing pulsatile information from all RGB channels, b)~the narrow ABP filter, and c)~the passband of ABP filter centered at precise HR frequency tracked by AMTC. To study the importance of each component, we conducted three controlled experiments by removing one factor at a time and the configurations of methods corresponding to the experiments are listed in Table~\ref{method configuration table}. The results for the methods are illustrated in Fig.~\ref{fig::barplot}. The height of each bin shows the average of correlation coefficient $\rho$ or the MAE of SpO$_2$ estimation results from testing sessions (SVR, PU case) of all participants. Each pair of error bars corresponds to the $95\%$ confidence interval that is calculated as $\pm1.96 \hat{\sigma} / \sqrt{N}$, where $\hat{\sigma}$ is the sample standard deviation and $N$ is the sample size/number of participants.

\begin{table}[!t]
\centering
\caption{\revm{Configurations for the ablation study of the proposed pipeline. The controlled experiments are conducted by replacing or removing one component at a time.}}
\label{method configuration table}
\begin{tabular}{c|ccc}
\hline
\multirow{3}{*}{Method Index} & \multicolumn{3}{c}{Configuration}             \\ \cline{2-4}  & \begin{tabular}[c]{@{}c@{}}Multi-channel \\ RoR features?\end{tabular} & \begin{tabular}[c]{@{}c@{}}Narrow \\ ABP filter?\end{tabular} & \begin{tabular}[c]{@{}c@{}}Accurate \\ HR tracking?\end{tabular} \\ \hline
I   & Two-channel RoR &\checkmark  &\checkmark (\scriptsize{AMTC})   \\\hline
II  & \checkmark & No ABP & n/a   \\\hline
III  & \checkmark & Wide ABP &\checkmark (\scriptsize{AMTC})  \\\hline
IV   &\checkmark  & \checkmark & Peak-finding  \\\hline
V  &\checkmark  & \checkmark & Weighted energy  \\\hline
Proposed   & \checkmark  & \checkmark  & \checkmark (\scriptsize{AMTC}) \\ \hline
\end{tabular}
\end{table}

\subsubsection{Advantage of The Proposed Multi-Channel RoR Over Two-Channel RoR} \label{subsubsec:abalation1}

In this part, we compare our proposed algorithm with ``\textbf{Method~(I): RoR with nABP (AMTC)}." Method~(I) follows the feature extraction method proposed in Section~\ref{subsec:feature-extraction}, including the \textit{narrow adaptive bandpass filter (nABP)} centered at AMTC-tracked HR. The only exception is that, instead of using the feature vector~$\mathbf{f}$ that contains multi-channel information, only the ratio of ratios between the red and blue channels as in traditional RoR methods is used. 

Fig.~\ref{fig::barplot} reveals that our proposed method outperforms method~(I) by a big margin. More specifically, our proposed method improves the correlation coefficient from $0.22$ to $0.68$ and the MAE from $1.67\%$ to $1.26\%$. This improvement confirms that our proposed multi-channel feature set helps with the more accurate SpO$_2$ monitoring.

\subsubsection{Contribution of Narrowband ABP Filter for Feature Extraction} 
\label{subsubsec:abalation2}
Here we compare with the following two methods to show the necessity of using a narrowband HR-guided bandpass filter:
\begin{itemize}
\item \textbf{Method~(II): Feature vector without ABP} uses a nonadaptive, generic bandpass filter with the passband over $[1, 2]$~Hz, covering the normal range of heart rate in sedentary mode to replace the HR-based narrow ABP filter proposed in Section~\ref{subsec:feature-extraction} for feature extraction.
\item \textbf{Method~(III): Feature vector with wide ABP (AMTC)} applies a wider ABP filter with $\pm0.5$~Hz bandwidth than the $\pm0.1$~Hz one used in our proposed method. This wider ABP filter's center frequency is provided by the AMTC tracking algorithm of the HR described in Section~\ref{subsec:hr-esti}.
\end{itemize}

\noindent The bandpass filters used for methods~(II)~and~(III) have the same bandwidth, $1$ Hz. In terms of center frequency, method~(II) used a fixed setting at $1.5$~Hz, while method~(III) is adaptively centered at the estimated HR value. Compared to method~(II), method~(III) has an improved testing MAE by $18\%$. Furthermore, compared to method~(III), our proposed method with a narrow ABP filter improves the correlation coefficient $\rho$ for testing by~$13\%$ and MAE by~$9\%$, suggesting the contribution of the narrow HR-based ABP filter strategy for AC computation.

\subsubsection{Importance of Accurate HR Tracking on SpO$_2$ Monitoring}\label{subsubsec:abalation3}
We consider the following two methods to compare with our proposed method:
\begin{itemize}
    \item \textbf{Method~(IV): Feature vector with narrow ABP (peak-finding)} applies a narrow ABP filter of bandwidth $\pm0.1$~Hz for extracting the feature vector $\mathbf{f}$. The center frequency of the ABP filter is the HR estimated from the peak-finding algorithm described in Section~\ref{subsec:hr-esti}. 
    \item \textbf{Method~(V): Feature vector with narrow ABP (weighted)} is similar to method~(IV), except that the frequency estimation algorithm is replaced by the weighted energy in Section~\ref{subsec:hr-esti}.
\end{itemize}
 
\noindent The averaged MAE of the HR estimation for all participants by the peak-finding algorithm, weighted frequency estimation algorithm, and AMTC algorithm are $7.11~(\pm3.66)$~bpm, $6.42~(\pm3.02)$~bpm, and $4.14~(\pm1.72)$~bpm, respectively.

Fig.~\ref{fig::barplot} shows that methods~(IV)~and~(V) perform similarly with $0.56$ vs. $0.57$ for correlation $\rho$ and $1.43\%$ vs. $1.40\%$ for MAE, respectively. Our proposed method guided by the AMTC tracked HR outperforms methods~(IV)~and~(V) by $21\%$ and $19\%$ in correlation, and by $12\%$ and $10\%$ in MAE, respectively. These results suggest that the accurate HR estimation for ABP filter design improves the quality of the AC magnitude by preserving the most cardiac-related signal from RGB channels, which in turn helps with accurate SpO$_2$ monitoring.

\subsection{Leave-One-Out Experiments}
\label{subsec:leaveoneout}

As a proof of concept and considering the currently limited amount of the available data, we have so far discussed the SpO$_2$ estimation under the \textit{participant-specific (PS)} scenario in Section~\ref{sec:expr} where the models are calibrated for each individual. This PS mode corresponds well to the trending ``precision telehealth" that tailors the healthcare service to individuals. 

In this subsection, we consider a more practical scenario where the test participant's data are never seen or only form a limited portion of the training data. In this scenario, we can develop a group-based model based on skin tone or other determinants of health, and for each subgroup, the model is ``universal" and participant-independent. We will examine this group-based model through the following two modes of leave-one-out experiments:

\begin{itemize}
\item \textit{Leave-one-session-out (LOSessO)}: when testing on a given participant, we include his/her training session data together with other participants' data for training.
\item \textit{Leave-one-participant-out (LOPartO)}: when testing on a given participant, we only use other people’s data for training and leave out the data from this test participant.
    
\end{itemize}


\begin{table}[!t]
\centering
\caption{Testing results of leave-one-participant-out (LOPartO) and leave-one-session-out (LOSessO) experiments, measured in the sample mean and the sample standard deviation (in parentheses).}
\label{leave one out table}
\begin{adjustbox}{width=0.49\textwidth}
\begin{tabular}{ccccccccc}
\hline
 & \multicolumn{2}{c}{LOPartO} &  & \multicolumn{2}{c}{LOSessO} &  & \multicolumn{2}{c}{PS} \\ \cline{2-3} \cline{5-6} \cline{8-9} 
& MAE & $\rho$ &  & MAE & $\rho$ &  & MAE & $\rho$ \\ \hline
\multirow{2}{*}{PU}
&1.70\%& 0.53 &    & 1.59\% & 0.55 &  &1.26\%  &0.68 \\
&($\pm$0.60\%)& ($\pm$0.38)  &  &($\pm$0.58\%)  & ($\pm$0.36) &  &($\pm$0.33\%)  &($\pm$0.10) \\ \hline
\multirow{2}{*}{PD} 
& 1.76\% & 0.48 &  & 1.70\% & 0.50 &  &  1.28\%&0.65 \\
& ($\pm$0.59\%) & ($\pm$0.38) &  & ($\pm$0.59\%) &($\pm$0.39)  &  &($\pm$0.40\%)  & ($\pm$0.14)\\ \hline
\end{tabular}
\end{adjustbox}
\end{table}

We group the participants by the skin type into lighter skin color (skin types II and III) and darker skin color (skin types IV and V) groups. We conduct LOSessO and LOPartO experiments on each subgroup and obtain the SVR generated testing results from all participants in Table~\ref{leave one out table}. The MAE and correlation coefficient $\rho$ improve from LOPartO to LOSessO to PS for both PU and PD cases. This result suggests that the precision telehealth inspired PS mode is the most accurate approach to monitoring SpO$_2$ for an individual. Based on the overall results shown in Table~\ref{leave one out table}, most participants demonstrate a consistent trend of the accuracy of SpO$_2$ estimation from LOPartO to LOSessO to PS case. The correlation $\rho$ of participant~\#12 is less than $-0.5$ in both leave-one-out modes, suggesting that this participant may have some uncommon relation compared to others between the extracted features and SpO$_2$ values.

\section{Discussion}

\subsection{Performance on Contact SpO$_2$ Monitoring}
\begin{table}[!t]
\centering
\caption{Comparison of the proposed algorithm in both contact and contact-free SpO$_2$ estimation settings.
The testing results are measured in the average MAE and correlation coefficient $\rho$.}
\label{table:contact-compare}
\begin{adjustbox}{width=0.49\textwidth}
\begin{tabular}{clccccc}
\hline
\multicolumn{1}{l}{}&& \multicolumn{2}{c}{\begin{tabular}[c]{@{}c@{}}Training \end{tabular}} & \multicolumn{1}{c}{} & \multicolumn{2}{c}{\begin{tabular}[c]{@{}c@{}} Testing \end{tabular}} \\ \cline{3-4} \cline{6-7} 
\multicolumn{1}{l}{}&& \multicolumn{1}{c}{MAE}&\multicolumn{1}{c}{$\rho$} & \multicolumn{1}{c}{}&\multicolumn{1}{c}{MAE}& \multicolumn{1}{c}{$\rho$} \\ \hline
\multicolumn{1}{c|}{\multirow{5}{*}{\begin{tabular}[c]{@{}c@{}}Contact \end{tabular}}} & RoR\cite{scully2011physiological} (LR)&1.60\%&0.54&&1.38\%&0.64
\\ \cline{2-7} 
\multicolumn{1}{c|}{}& RoR\cite{scully2011physiological} (SVR)&1.14\%&0.73&&1.32\%&0.60
\\ \cline{2-7} 
\multicolumn{1}{c|}{}& RoR\cite{lu2015prototype} (LR) &1.47\%&0.62&&1.39\%&0.63
\\ \cline{2-7} 
\multicolumn{1}{c|}{}& RoR\cite{lu2015prototype} (SVR) &0.99\%&0.83&&1.27\%&0.66
\\ \cline{2-7} 
\multicolumn{1}{c|}{} & \textbf{Proposed} &\textbf{0.91\%}&\textbf{0.84}&&\textbf{1.17\%}&\textbf{0.81}
\\ \hline

\multicolumn{1}{c|}{\multirow{2}{*}{\begin{tabular}[c]{@{}c@{}}Contact-free \end{tabular}}} & RoR (2-channel) &1.61\%&\textbf{0.73}&&1.75\%&0.36
\\ \cline{2-7} 
\multicolumn{1}{c|}{}& \textbf{Proposed}&\textbf{1.36\%}&0.62&&\textbf{1.29\%}&\textbf{0.68}
\\ \hline

\end{tabular}
\end{adjustbox}
\end{table}

In addition to contact-free SpO$_2$ monitoring, we evaluate whether our proposed algorithm can be applied to a contact-based smartphone setup. To collect data, the left index finger covers the smartphone's illuminating flashlight and the nearby built-in camera, and the camera captures a pulse video at the fingertip. Another smartphone is used to simultaneously record a top view video of the back side of the right hand whose index finger is placed in the oximeter for SpO$_2$ reference data collection. One participant took part in this extended experiment where one training session with three breath-holding cycles was recorded, and three testing sessions were recorded 30 minutes after the training session.

In Table~\ref{table:contact-compare}, we compare the performance of our proposed algorithm in both the contact-based and contact-free SpO$_2$ measurement settings. The conventional RoR models used in~\cite{scully2011physiological} and~\cite{lu2015prototype} were implemented as baseline models for contact-based SpO$_2$ measurement. In~\cite{scully2011physiological}, the mean and standard deviation of each window from the red and blue channels are calculated as the DC and AC components. A linear model was built to relate the ratio-of-ratios from the two color channels with SpO$_2$. In~\cite{lu2015prototype}, the median of the pulsatile peak-to-valley amplitude is regarded as the AC component. For the two RoR methods, we implemented both LR and SVR. For contact-free SpO$_2$ measurement, we take the traditional two-color channel RoR method implemented in Section~\ref{subsec:ablation-study} as the baseline to compare with the proposed method.

Table~\ref{table:contact-compare} reveals that our proposed algorithm outperforms other conventional RoR models in the contact-based SpO$_2$ monitoring. Even in the contact-free case, our algorithm presents a comparable performance to that of the contact-based cases, despite that the SNR of the fingertip video is better than the SNR from a remote hand video.

\subsection{Resilience Against Blurring}

\begin{figure}[!t]
\centering
\includegraphics[width=3.5in]{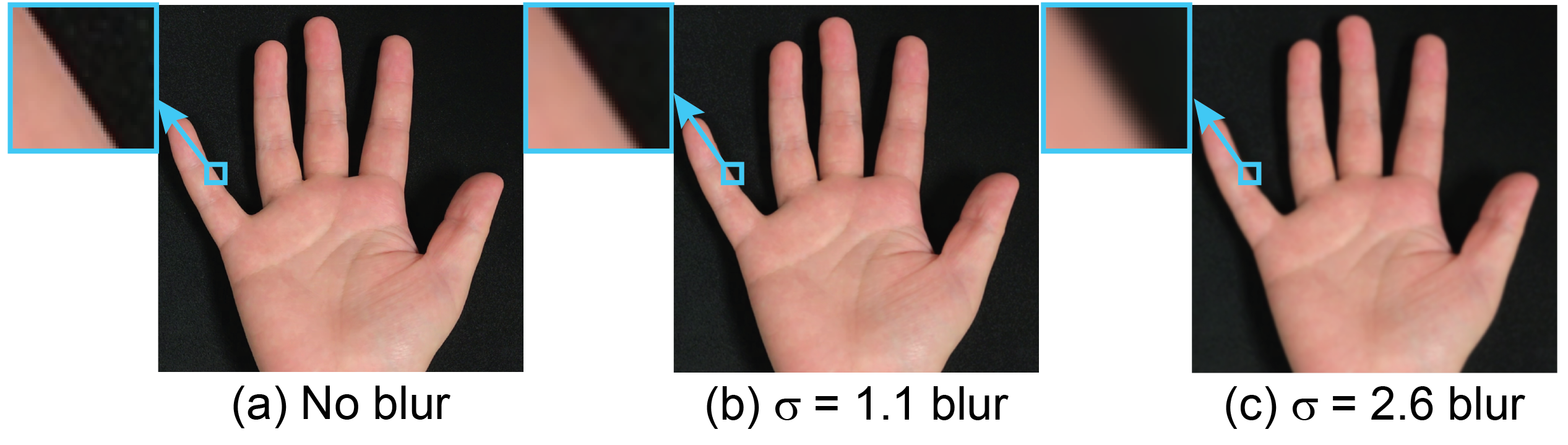}
\caption{\revm{Illustration of blurring effects using different blurry levels $\sigma$ on hand videos. The wider the kernel is, the blurrier the videos are.}}
\label{fig::blurring}
\end{figure} 

In this subsection, we explore the robustness of our algorithm to the blurring effect on hand images. In the current setup, the hands are placed on a stable table with a cellphone camera acquiring the skin color of both hands. Ideal laboratory conditions are often not satisfied under practical scenarios, and the hand images captured by the cellphone cameras may be blurred due to being out of focus. The point spread function is modeled as a 2D homogeneous Gaussian kernel. 
The finite support of the kernel is defined manually to generate perceptually different blurry effects and then the standard deviation $\sigma$ is computed based on the given support. To test different blurry effects, we experimented with two different blurry levels $\sigma = 1.1$~($5 \times 5$ pixels) and $\sigma=2.6$~($15 \times 15$ pixels), respectively. \revm{We show the blurring effects in Fig.~\ref{fig::blurring}.}

\begin{table}[!t]
\centering
\caption{Simulation for Gaussian blurring effect on hand videos. SVR generated results for PU cases are listed for different $\sigma$ and Gaussian kernel sizes. The results are quantified in terms of the sample mean and sample standard deviation (in parentheses).}
\label{blur effect}
\begin{adjustbox}{width=0.49\textwidth}

\begin{tabular}{cccccc}
\hline & \multicolumn{2}{c}{Training}&&\multicolumn{2}{c}{Testing}\\ \cline{2-3} \cline{5-6}& MAE & $\rho$&& MAE & $\rho$ \\ \hline
$\sigma=2.6$ blur  &1.41\%&0.72&&1.31\%&0.67 \\ 
($15 \times 15$ pixels) &($\pm$0.50\%)&($\pm$0.11)&&($\pm$0.35\%)&($\pm$0.09)\\ \hline

$\sigma=1.1$ blur&1.42\%&0.70&&1.34\%&0.68\\ 
($5 \times 5$ pixels)&($\pm$0.59\%)&($\pm$0.16)&&($\pm$0.41\%)&($\pm$0.10)\\ \hline

\multirow{2}{*}{No blur}&1.33\%&0.76&&1.26\%&0.68\\ 
&($\pm$0.54\%)&($\pm$0.09)&&($\pm$0.33\%)&($\pm$0.10)\\ \hline

\end{tabular}
\end{adjustbox}
\end{table}

Table~\ref{blur effect} presents the SVR generated results for PU cases with different $\sigma$ and kernel sizes. We use the SVR, PU scenario to showcase here as it achieves the best SpO$_2$ prediction performance, which is verified in Section~\ref{subsec:results-main}. From the table, we find that our algorithm is robust to the Gaussian blurring effect. After the $\sigma = 1.1$ blurring, the testing $\rho$ remains the same, and testing MAE is $6.3\%$ higher than the no blurring case. After the $\sigma=2.6$ blurring, the testing $\rho$ is $1.5\%$ lower and MAE is $4.0\%$ higher than the no blurring case.

\subsection{Limitations and Further Verification with Different Hypoxia Protocols}

From the recordings of our data collection protocol for voluntary breath-holding, we observed that HR and SpO$_2$ are correlated for many participants. That is, in one breath-holding cycle, when the participant starts to hold breath, his/her HR increases and SpO$_2$ drops as the oxygen runs out. As he/she resumes normal breathing, his/her HR and SpO$_2$ recover to be within the normal range. Due to individuals' different physical conditions, in some participants, the peak of the HR signal and valley of the SpO$_2$ signal happen in such a short time interval that HR and SpO$_2$ are significantly negatively correlated. This observation is in line with the biological literature~\cite{grunovas2016cardiovascular}. In the literature, breath-holding exercises were found to be able to yield significant changes in the cardiovascular system. In the central circulation, they caused significant changes in heart rate, and in the peripheral circulation, they caused significant changes in arterial blood flow and oxygen saturation.

Based on the above observation that HR is correlated with SpO$_2$ during breath-holding, we are curious whether our method also works for a different protocol where the HR remains relatively constant compared to SpO$_2$. An \emph{intermittent hypoxia protocol} used in the literature shows that by receiving hypoxic air (inspired fraction of oxygen between $12\%$ and $15\%$) intermittently with normoxic air, the participant can have a much milder HR change than breath-holding, while a significant decrease in SpO$_2$ can be achieved during the hypoxia~\cite{faulhaber2015heart, koehle2008effect}. The research restriction affecting human subject research in many U.S. institutions limited our ability to carry out the abovementioned hypoxia protocol. Once the restriction is eased, we will investigate the performance of our proposed algorithm when applied to other hypoxia protocols.

\section{Conclusion and Future Work}
This paper presents a contact-free method of measuring blood oxygen saturation from hand videos captured by smartphone cameras. Our proposed method is a synergistic combination of several key components, including the multi-channel ratio-of-ratios feature set, the narrowband filtering that adaptively centered at heart rates, and the accurately estimated heart rate. We have seen encouraging results of a mean absolute error of $1.26\%$ with a commercial pulse oximeter as the reference, outperforming the conventional ratio-of-ratios method by $25\%$. We have also analyzed the impact of sides of the hand and skin tones on the SpO$_2$ estimation. We have found that, given our collected dataset, the palm side performs well regardless of the skin tone; for palm-up cases, we do not observe significant performance differences between lighter and darker skin tones.

Future work includes verifying our methods with data collected under different hypoxia protocols, enlarging the dataset with more variations in skin color, and applying large-scale and interpretable neural networks with optophysiological insights. \rev{Future development may also consider real-life scenarios wherein participants walk toward the camera or have other substantial movement and investigate how to achieve good SpO$_2$ estimation from video in these cases.}

\balance
\bibliographystyle{IEEEtran}
\bibliography{refs}

\begin{thebibliography}{10}
\providecommand{\url}[1]{#1}
\csname url@samestyle\endcsname
\providecommand{\newblock}{\relax}
\providecommand{\bibinfo}[2]{#2}
\providecommand{\BIBentrySTDinterwordspacing}{\spaceskip=0pt\relax}
\providecommand{\BIBentryALTinterwordstretchfactor}{4}
\providecommand{\BIBentryALTinterwordspacing}{\spaceskip=\fontdimen2\font plus
\BIBentryALTinterwordstretchfactor\fontdimen3\font minus
  \fontdimen4\font\relax}
\providecommand{\BIBforeignlanguage}[2]{{%
\expandafter\ifx\csname l@#1\endcsname\relax
\typeout{** WARNING: IEEEtran.bst: No hyphenation pattern has been}%
\typeout{** loaded for the language `#1'. Using the pattern for}%
\typeout{** the default language instead.}%
\else
\language=\csname l@#1\endcsname
\fi
#2}}
\providecommand{\BIBdecl}{\relax}
\BIBdecl

\bibitem{simon2008role}
M.~C. Simon and B.~Keith, ``{The role of oxygen availability in embryonic
  development and stem cell function},'' \emph{Nature Reviews Molecular Cell
  Biology}, vol.~9, no.~4, pp. 285--296, Apr. 2008.

\bibitem{shenoy2020considerations}
N.~Shenoy, R.~Luchtel, and P.~Gulani, ``{Considerations for target oxygen
  saturation in COVID-19 patients: Are we under-shooting?}'' \emph{BMC
  Medicine}, vol.~18, no.~1, pp. 1--6, Dec. 2020.

\bibitem{Couzin-Frankel455}
\BIBentryALTinterwordspacing
J.~Couzin-Frankel, ``{The mystery of the pandemic{\textquoteright}s
  {\textquoteleft}happy hypoxia{\textquoteright}},'' \emph{Science}, 2020.
  [Online]. Available:
  \url{https://science.sciencemag.org/content/368/6490/455}
\BIBentrySTDinterwordspacing

\bibitem{tobin2020covid}
M.~J. Tobin, F.~Laghi, and A.~Jubran, ``{Why COVID-19 silent hypoxemia is
  baffling to physicians},'' \emph{American Journal of Respiratory and Critical
  Care Medicine}, vol. 202, no.~3, pp. 356--360, Aug. 2020.

\bibitem{teo2020early}
J.~Teo, ``{Early detection of silent hypoxia in COVID-19 pneumonia using
  smartphone pulse oximetry},'' \emph{J. Medical Systems}, vol.~44, no.~8, pp.
  1--2, Aug. 2020.

\bibitem{webster1997design}
J.~G. Webster, \emph{{Design of pulse oximeters}}.\hskip 1em plus 0.5em minus
  0.4em\relax CRC Press, Oct. 1997.

\bibitem{severinghaus2007takuo}
J.~W. Severinghaus, ``{Takuo Aoyagi: Discovery of pulse oximetry},''
  \emph{Anesthesia and Analgesia}, vol. 105, no.~6, pp. S1--S4, Dec. 2007.

\bibitem{pluddemann2011pulse}
A.~Pl{\"u}ddemann, M.~Thompson, C.~Heneghan, and C.~Price, ``{Pulse oximetry in
  primary care: Primary care diagnostic technology update},'' \emph{British J.
  General Practice}, vol.~61, no. 586, pp. 358--359, May 2011.

\bibitem{applespo2}
``{How to use the Blood Oxygen app on Apple Watch Series 6},''
  \url{https://support.apple.com/en-us/HT211027}, {Accessed: 2021-05-17}.

\bibitem{scully2011physiological}
C.~G. Scully, J.~Lee, J.~Meyer, A.~M. Gorbach, D.~Granquist-Fraser,
  Y.~Mendelson, and K.~H. Chon, ``{Physiological parameter monitoring from
  optical recordings with a mobile phone},'' \emph{IEEE Trans. Biomed. Eng.},
  vol.~59, no.~2, pp. 303--306, Jul. 2011.

\bibitem{lu2015prototype}
Z.~Lu, X.~Chen, Z.~Dong, Z.~Zhao, and X.~Zhang, ``{A prototype of reflection
  pulse oximeter designed for mobile healthcare},'' \emph{IEEE J. Biomed.
  Health Inform.}, vol.~20, no.~5, pp. 1309--1320, Aug. 2015.

\bibitem{ding2018measuring}
X.~Ding, D.~Nassehi, and E.~C. Larson, ``{Measuring oxygen saturation with
  smartphone cameras using convolutional neural networks},'' \emph{IEEE J.
  Biomed. Health Inform.}, vol.~23, no.~6, pp. 2603--2610, Dec. 2018.

\bibitem{bui2020smartphone}
N.~Bui, A.~Nguyen, P.~Nguyen, H.~Truong, A.~Ashok, T.~Dinh, R.~Deterding, and
  T.~Vu, ``{Smartphone-based SpO$_2$ measurement by exploiting wavelengths
  separation and chromophore compensation},'' \emph{ACM Trans. Sens. Netw.},
  vol.~16, no.~1, pp. 1--30, Jan. 2020.

\bibitem{tayfur2019reliability}
{\.I}.~Tayfur and M.~A. Afacan, ``{Reliability of smartphone measurements of
  vital parameters: A prospective study using a reference method},'' \emph{The
  American J. Emergency Medicine}, vol.~37, no.~8, pp. 1527--1530, Aug. 2019.

\bibitem{kong2013non}
L.~Kong, Y.~Zhao, L.~Dong, Y.~Jian, X.~Jin, B.~Li, Y.~Feng, M.~Liu, X.~Liu, and
  H.~Wu, ``{Non-contact detection of oxygen saturation based on visible light
  imaging device using ambient light},'' \emph{Opt. Exp.}, vol.~21, no.~15, pp.
  17\,464--17\,471, Jul. 2013.

\bibitem{van2016new}
M.~Van~Gastel, S.~Stuijk, and G.~de~Haan, ``{New principle for measuring
  arterial blood oxygenation, enabling motion-robust remote monitoring},''
  \emph{Scientific Reports}, vol.~6, no.~1, pp. 1--16, Dec. 2016.

\bibitem{guazzi2015non}
A.~R. Guazzi, M.~Villarroel, J.~Jorge, J.~Daly, M.~C. Frise, P.~A. Robbins, and
  L.~Tarassenko, ``{Non-contact measurement of oxygen saturation with an RGB
  camera},'' \emph{Biomed. Opt. Express}, vol.~6, no.~9, pp. 3320--3338, Sep.
  2015.

\bibitem{van2019data}
M.~van Gastel, W.~Verkruysse, and G.~de~Haan, ``{Data-driven calibration
  estimation for robust remote pulse-oximetry},'' \emph{Applied Sciences},
  vol.~9, no.~18, p. 3857, Jan. 2019.

\bibitem{shao2015noncontact}
D.~Shao, C.~Liu, F.~Tsow, Y.~Yang, Z.~Du, R.~Iriya, H.~Yu, and N.~Tao,
  ``{Noncontact monitoring of blood oxygen saturation using camera and
  dual-wavelength imaging system},'' \emph{IEEE Trans. Biomed. Eng.}, vol.~63,
  no.~6, pp. 1091--1098, Sep. 2015.

\bibitem{tsai2016no}
H.-Y. Tsai, K.-C. Huang, and J.~A. Yeh, ``{No-contact oxygen saturation
  measuring technology for skin tissue and its application},'' \emph{IEEE
  Instrum. Meas. Magazine}, vol.~19, no.~5, pp. 57--64, Sep. 2016.

\bibitem{cote1988effect}
C.~J. Cot{\'e}, E.~A. Goldstein, W.~H. Fuchsman, and D.~C. Hoaglin, ``{The
  effect of nail polish on pulse oximetry},'' \emph{Anesthesia and Analgesia},
  vol.~67, no.~7, pp. 683--686, Jul. 1988.

\bibitem{yont2014effect}
G.~H. Y{\"o}nt, E.~A. Korhan, and B.~Dizer, ``{The effect of nail polish on
  pulse oximetry readings},'' \emph{Intensive and Critical Care Nursing},
  vol.~30, no.~2, pp. 111--115, Apr. 2014.

\bibitem{rosa2019noncontact}
A.~d. F.~G. Rosa and R.~C. Betini, ``{Noncontact SpO$_2$ measurement using
  Eulerian video magnification},'' \emph{IEEE Trans. Instrum. Meas.}, vol.~69,
  no.~5, pp. 2120--2130, May 2019.

\bibitem{casalino2020mhealth}
G.~{Casalino}, G.~{Castellano}, and G.~{Zaza}, ``{A mHealth solution for
  contact-less self-monitoring of blood oxygen saturation},'' in \emph{{IEEE
  Symp. Comp. Comm.}}, Jul. 2020, pp. 1--7.

\bibitem{bal2015non}
U.~Bal, ``{Non-contact estimation of heart rate and oxygen saturation using
  ambient light},'' \emph{Biomed. Opt. Exp.}, vol.~6, no.~1, pp. 86--97, Jan.
  2015.

\bibitem{tarassenko2014non}
L.~Tarassenko, M.~Villarroel, A.~Guazzi, J.~Jorge, D.~Clifton, and C.~Pugh,
  ``{Non-contact video-based vital sign monitoring using ambient light and
  auto-regressive models},'' \emph{Physiol. Meas}, vol.~35, no.~5, p. 807, Mar.
  2014.

\bibitem{sun2021robust}
Z.~Sun, Q.~He, Y.~Li, W.~Wang, and R.~K. Wang, ``{Robust non-contact peripheral
  oxygenation saturation measurement using smartphone-enabled imaging
  photoplethysmography},'' \emph{Biomed. Opt. Exp.}, vol.~12, no.~3, pp.
  1746--1760, Mar. 2021.

\bibitem{sun2015photoplethysmography}
Y.~Sun and N.~Thakor, ``{Photoplethysmography revisited: From contact to
  noncontact, from point to imaging},'' \emph{IEEE Trans. Biomed. Eng.},
  vol.~63, no.~3, pp. 463--477, 2015.

\bibitem{wang2016algorithmic}
W.~Wang, A.~C. den Brinker, S.~Stuijk, and G.~de~Haan, ``{Algorithmic
  principles of remote PPG},'' \emph{IEEE Trans. Biomed. Eng.}, vol.~64, no.~7,
  pp. 1479--1491, Sep. 2016.

\bibitem{szeliski2010computer}
R.~Szeliski, \emph{{Computer Vision: Algorithms and Applications}}.\hskip 1em
  plus 0.5em minus 0.4em\relax Springer Science \& Business Media, 2010.

\bibitem{urooj2018analysis}
A.~Urooj and A.~Borji, ``Analysis of hand segmentation in the wild,'' in
  \emph{Proceedings of the IEEE Conference on Computer Vision and Pattern
  Recognition (CVPR)}, 2018, pp. 4710--4719.

\bibitem{chai1999face}
D.~Chai and K.~N. Ngan, ``{Face segmentation using skin-color map in videophone
  applications},'' \emph{IEEE Trans. Circuits and Systems for Video
  Technology}, vol.~9, no.~4, pp. 551--564, Jun. 1999.

\bibitem{otsu1979threshold}
N.~Otsu, ``A threshold selection method from gray-level histograms,''
  \emph{IEEE Trans. Syst., Man, and Cybernet.}, vol.~9, no.~1, pp. 62--66, Jan.
  1979.

\bibitem{zhu2020adaptive}
Q.~Zhu, M.~Chen, C.-W. Wong, and M.~Wu, ``{Adaptive multi-trace carving for
  robust frequency tracking in forensic applications},'' \emph{IEEE Trans. Inf.
  Forensics Security}, vol.~16, pp. 1174--1189, May 2020.

\bibitem{hajj2012instantaneous}
A.~Hajj-Ahmad, R.~Garg, and M.~Wu, ``{Instantaneous frequency estimation and
  localization for ENF signals},'' in \emph{Proc. 4th Annu. Summit and Conf.
  (APSIPA)}, Dec. 2012, pp. 1--10.

\bibitem{chang2011libsvm}
C.-C. Chang and C.-J. Lin, ``{LIBSVM: A library for support vector machines},''
  \emph{ACM Trans. Intelligent Systems and Technology}, vol.~2, no.~3, pp.
  1--27, May 2011.

\bibitem{theodoridis2015machine}
S.~Theodoridis, \emph{Machine Learning: A Bayesian and Optimization
  Perspective}.\hskip 1em plus 0.5em minus 0.4em\relax Academic press, 2015.

\bibitem{skin-types}
\BIBentryALTinterwordspacing
\emph{Fitzpatrick skin phototype}, Australian Radiation Protection and Nuclear
  Safety Agency, accessed: 2022-01-16. [Online]. Available:
  \url{https://www.arpansa.gov.au/sites/default/files/legacy/pubs/RadiationProtection/FitzpatrickSkinType.pdf}
\BIBentrySTDinterwordspacing

\bibitem{grunovas2016cardiovascular}
A.~Grunovas, E.~Trinkunas, A.~Buliuolis, E.~Venskaityte, and J.~Poderys,
  ``{Cardiovascular response to breath-holding explained by changes of the
  indices and their dynamic interactions},'' \emph{Biological Systems: Open
  Access}, vol.~5, p. 152, 2016.

\bibitem{faulhaber2015heart}
M.~Faulhaber, H.~Gatterer, T.~Haider, T.~Linser, N.~Netzer, and M.~Burtscher,
  ``{Heart rate and blood pressure responses during hypoxic cycles of a 3-week
  intermittent hypoxia breathing program in patients at risk for or with mild
  COPD},'' \emph{Int. J. Chronic Obstructive Pulmonary Disease}, vol.~10, p.
  339, 2015.

\bibitem{koehle2008effect}
M.~Koehle, W.~Sheel, W.~Milsom, and D.~McKenzie, ``{The effect of two different
  intermittent hypoxia protocols on ventilatory responses to hypoxia and carbon
  dioxide at rest},'' in \emph{Integration in Respiratory Control}.\hskip 1em
  plus 0.5em minus 0.4em\relax Springer, 2008, pp. 218--223.

\end{thebibliography}


\vspace{-10mm}
\begin{IEEEbiography}[{\includegraphics[width=1in,height=1.25in,clip,keepaspectratio]{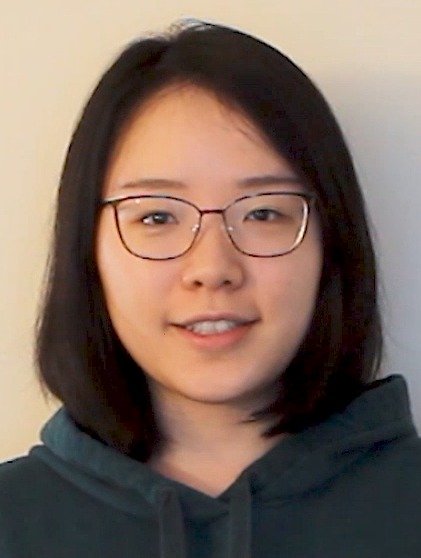}}]{Xin Tian} (Graduate Student Member, IEEE) received the B.E. degree in Optoelectronic Information Science and Engineering from Huazhong University of Science and Technology of China in 2017. She is currently pursuing a Ph.D. degree at the Department of Electrical and Computer Engineering, University of Maryland, College Park. Her current research interests are signal processing, data science, and machine learning in smart health. She was a summer intern at Facebook in 2020. She was selected as a Future Faculty Fellow by the University of Maryland Clark School of Engineering in 2020. 
\end{IEEEbiography}

\vspace{-10mm}
\begin{IEEEbiography}[{\includegraphics[width=1in,height=1.25in,clip,keepaspectratio]{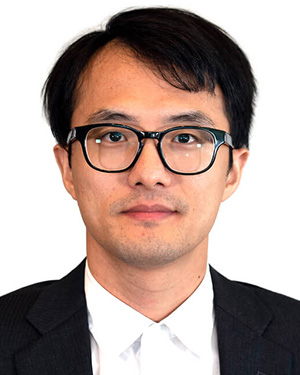}}]{Chau-Wai Wong}
(Member, IEEE) received his B.Eng. and M.Phil. degrees in electronic and information engineering from The Hong Kong Polytechnic University, in 2008 and 2010, and the Ph.D. degree in electrical engineering from the University of Maryland, College Park, MD, USA, in 2017. He is currently an Assistant Professor with the Department of Electrical and Computer Engineering, the Forensic Sciences Cluster, and the Secure Computing Institute, North Carolina State University. He was a data scientist at Origin Wireless, Inc., Greenbelt, MD, USA. His research interests include multimedia forensics, statistical signal processing, machine learning, data analytics, and video coding. Dr. Wong received a Top-Four Student Paper Award, Future Faculty Fellowship, HSBC Scholarship, and Hitachi Scholarship. He was the General Secretary of the IEEE PolyU Student Branch from 2006 to 2007. He was involved in organizing the third edition of the IEEE Signal Processing Cup in 2016 on electric network frequency forensics.
\end{IEEEbiography}

\vspace{-10mm}
\begin{IEEEbiography}[{\includegraphics[width=1in,height=1.25in,clip,keepaspectratio]{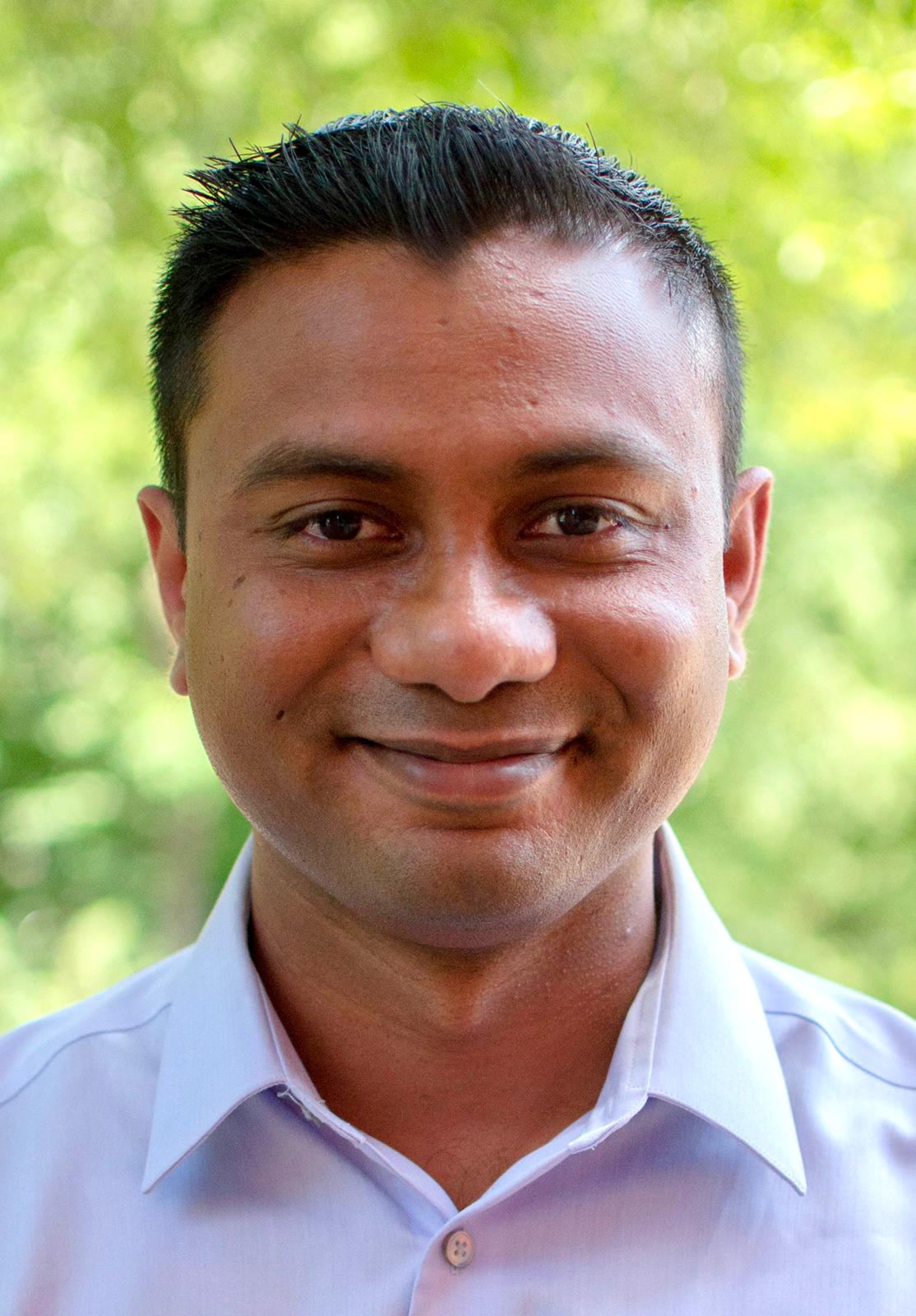}}]{Sushant~M.~Ranadive} received his bachelor’s degree in Occupational Therapy from India, followed by a clinical master’s in Exercise Physiology at the University of Illinois at Chicago and {a} Ph.D. in Kinesiology from the University of Illinois at Urbana-Champaign. He was a postdoctoral fellow and research faculty member at the Mayo Clinic before joining the University of Maryland as an Assistant Professor in Kinesiology in 2017. His research interests include studying the mechanisms related to the cardiovascular health of women as compared with age-matched men with social determinants of health as a factor, and racial and sex disparities in vascular function following acute inflammation. He has received NSF and NIH funding as a co-investigator and research awards at the University of Maryland. He has had major committee appointments at the American College of Sports Medicine Mid-Atlantic Chapter and the American Physiological Society.

\end{IEEEbiography}

\vspace{-10mm}
\begin{IEEEbiography}[{\includegraphics[width=1in,height=1.25in,clip,keepaspectratio]{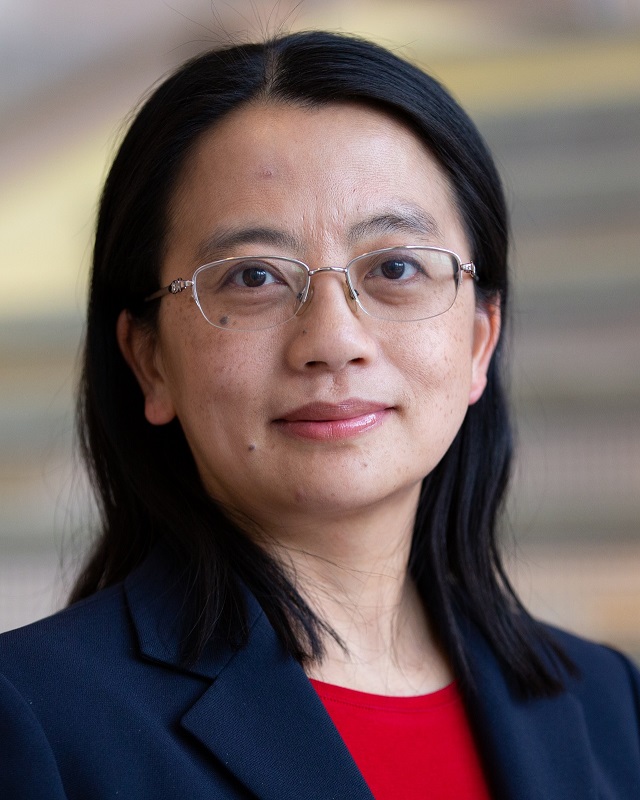}}]{Min Wu} (Fellow, IEEE) received the B.E. degree in automation and the B.A. degree in economics from Tsinghua University, Beijing, China, in 1996 with the highest honors, and the Ph.D. degree in electrical engineering from Princeton University, Princeton, NJ, USA, in 2001. She is a Professor and an Associate Dean of Engineering and a Distinguished Scholar-Teacher with the University of Maryland, College Park, MD, USA. Her research interests include information security and forensics, multimedia signal processing, and applications of data science and machine learning in health and IoT.  Prof. Wu received a U.S. NSF CAREER award, a U.S. ONR Young Investigator Award, a TR100 Young Innovator Award from the MIT Technology Review, an IEEE Harriett B. Rigas Education Award, and an IEEE SP Society Meritorious Service Award. She chaired the IEEE Technical Committee on Information Forensics and Security, and has served as the Vice President—Finance of the IEEE Signal Processing Society and the Editor-in-Chief of the IEEE Signal Processing Magazine. She has been elected as 2022-2023 President-Elect of the IEEE Signal Processing Society.  She is a Fellow of AAAS and of the U.S. National Academy of Inventors.

\end{IEEEbiography}

\end{document}